\documentstyle[sprocl]{article}

\def\be{\begin{equation}}
\def\ee{\end{equation}}
\def\bea{\begin{eqnarray}}
\def\eea{\end{eqnarray}}

\def\gtrsim{\mathrel{\hbox{\rlap{\hbox{\lower4pt\hbox{$\sim$}}}\hbox{$>$}}}}
\def\kms{\mbox{$\ {\rm km\ s}^{-1}$}}
\def\kpc{\mbox{$\ {\rm kpc}$}}
\def\lesssim{\mathrel{\hbox{\rlap{\hbox{\lower4pt\hbox{$\sim$}}}\hbox{$<$}}}}
\def\lya{\mbox{${\rm Ly}\alpha$}}
\def\mpc{\mbox{$\ {\rm Mpc}$}}
\def\msol{\mbox{$\ {\rm M}_\odot$}}

\let\la=\lesssim			

\begin{document}

\title{Concluding Remarks}

\author{A. Yahil}

\address{Department of Physics and Astronomy, State University of New York,\\
Stony Brook, NY 11794-3800, USA}


\maketitle

\section{Introduction}

Research in cosmology traditionally divided into two separate lines.  On the
one hand was the search for initial conditions: the cosmological parameters
$H$, $\Omega$, $\Omega_b$, and $\Lambda$, and the power spectrum $P(k)$.  On
the other hand was the study of the formation and evolution of structures, from
globular-cluster sized objects to large-scale structures.  These lines of
investigation are now becoming increasingly intertwined, and this tendency
manifested itself in this conference.  Following is my personal perspective on
where we stand today.  I also discuss some technological developments that I
think will affect the field.  My concluding remarks are very informal.  They
are not a review paper, and I omit references altogether rather than give a
partial list.

\section{Cosmological Density Parameter}

The preponderance of the evidence until 1985 was for an open universe, $\Omega
\la 0.2$.  Then came the observations of large-scale flows, particularly when
directly compared with the density structures measured by IRAS.  Together they
have favored $\Omega=1$, the value predicted by the popular theory of
inflation.

Now the pendulum has swung back in favor of an open universe.  The main lines
of evidence consist of (1)~virial analyses of clusters of galaxies, (2)~the
ratio of baryonic to dark matter in the intracluster medium, when combined with
the predictions of Big-Bang nucleosynthesis, (3)~direct estimates of cluster
masses from the analysis of lensed arcs, (4)~the low galaxian velocity
dispersion in the Local Supercluster, and, in recent studies, (5)~the evolution
of the abundance of clusters of galaxies, and (6)~the magnitude-redshift
relation of type-I supernovae.

But there is still evidence for higher $\Omega$.  The most recent IRAS--POTENT
comparison, while reducing somewhat its estimate of $\Omega$ is still
inconsistent with values as low as $\Omega\approx 0.2-0.3$.  I am also worried
about the strong decrease in the abundance of galaxies at high redshifts, which
is progressively more difficult to understand in high-volume universes, i.e.,
open universes, or, worse yet, $\lambda$ models.

The question will be settled very conclusively in a few years when the MAP and
Planck satellites are launched and measure the CBR anisotropy with sufficient
precision to allow the determination of all the cosmological parameters to an
accuracy of a few percent.  (To be on the safe side, I am discounting some of
the stronger claims to accuracy.)

\section{Power Spectrum}

The power spectrum is today well measured on the scale of COBE and on the scale
of galaxy clustering.  The interpretation of both measurements is subject to
uncertainty, the COBE anisotropy because it may be partly due to tensor
fluctuations that do not grow, and clustering on the scale of galaxies because
their formation may be biased.

The situation is even murkier at intermediate scales, $k \sim {\rm few} \times
0.01 \mpc$, for which existent measurements of the CBR anisotropy and the
galaxy correlation function are both inaccurate and contradictory.  Nor am I
swayed by the recent attempt to measure the power spectrum from the unsaturated
\lya\ forest.  I worry that, because the \lya\ clouds are so highly ionized,
the residual density of neutral hydrogen that provides the absorption is very
sensitive to the ionizing radiation.  Spatial fluctuations in the ionizing
radiation therefore also create correlations in the \lya\ forest.  The
situation is even worse when we consider the evolution of the ionizing
radiation with time, which is not well known at all.

One method that holds great promise of determining the mass power spectrum is
weak random-field lensing by large-scale structure.  This method requires
careful imaging over large areas and is now underway using mosaics of CCDs.  We
should hear first results soon.

Soon to come are also the wide-angle galaxy surveys of the 2DF and Sloane that
promise to nail the galaxy power spectrum on large scales, using $\sim 10^6$
galaxies.  An alternative that Ana Campos and I are pursuing is to measure the
angular power spectrum in deep images of a smaller region of the sky.  The
increased depth results in comparable numbers of galaxies and spatial
separations.  An added advantage over the wide-angle surveys is the measurement
of the evolution of the power spectrum, since the median redshift of the
galaxies approaches $z=1$.

\section{Structure Formation and Evolution}

The important advance of the past three years has been the direct observation
of high-redshift galaxies.  Hypotheses of galaxy formation at high redshift can
now be tested directly at or close to the redshift at which the galaxies are
supposed to form.

The main conclusion from deep imaging observations to date, particularly those
of the Hubble Deep Field, is that high-redshift galaxies are small, with
characteristic sizes $\sim 1\kpc$, not $\sim 10\kpc$.  Since the galaxies are
observed at rest UV wavelengths, and since the UV emission in nearby galaxies
is also confined to starbursting regions of comparable size, it is not yet
certain whether high-redshift galaxies are indeed small, destined to grow with
time by mergers and/or accretion, or whether we just observe the parts of the
galaxies that happen to starburst.  Preliminary work suggests that
high-redshift galaxies may indeed be smaller than present-day galaxies, because
larger galaxies could be seen.  However, the interpretation of the
observational limits is sensitive to uncertainties in the evolution of galaxies
and in the amount and distribution of dust in them.

There are now some limits on the epoch of star formation in elliptical galaxies
based on observations of such galaxies at redshifts $z\approx 1-1.5$, which
show no sign of any significant star formation for $\sim 2$ Gyr prior to the
epoch at which the galaxies are observed.  However, that does not mean that the
galaxies had their observed sizes throughout that period.  Star formation could
have taken place in smaller sub-units which later merged.

It is important to note that the deep images observe only the stellar content
of galaxies (and UV-emitting stars at that).  Recent observations of damped
\lya\ lines by Prochaska and Wolfe suggest the existence of large, rotating,
gas disks, whose sizes are, in fact, even larger than present-day galaxies.

Observations of nearby spirals also suggest that the gas can last a significant
time before it turns into stars.  Judy Young and collaborators have shown that
Sa and even some S0 galaxies do not necessarily have less gas than Sb and Sc
galaxies.  They simply do not produce massive stars as rapidly.  Furthermore,
stars in disks may be forming from the inside out.  Kennicutt has shown that
the star-formation rate is not constant as a function of galactocentric radius.
A better characterization is that a fixed fraction of the gas, $\sim 5\%$,
forms stars per orbit.  Since galaxies have essentially flat rotation curves,
it follows that the rate of star formation is approximately inversely
proportional to radius.

It should be added parenthetically that, if stellar disks indeed grow from the
inside out, then white dwarfs could offer a more stringent age limit than
globular clusters.  Normally we look for the oldest stars to set the limit, but
if we can convincingly argue that the white dwarfs in the solar neighborhood
formed later, say at redshifts $z\la 1$, then the current maximum white dwarf
ages of 10 Gyr suddenly become very interesting.

All in all, the directly observed UV radiation from high-redshift galaxies
corresponds to a star-formation rate of only $\sim 10\msol/{\rm yr}$, less than
the 100\msol/yr required to make $10^{11}\msol$ in 1 Gyr.  A plausible
explanation of the deficit would be a factor $\sim 10$ absorption by dust.
There are a number of theoretical estimates of the dust content of
high-redshift galaxies, but no observations.  The classical dust measure is the
ratio of Balmer lines, which for objects at redshifts $z\approx 1-1.5$ can be
observed in the near infrared.

\section{\boldmath $N$-Body and Hydrodynamic Simulations}

Large $N$-Body simulations have now been around for over a decade and are
beginning to become real tools for determining cosmological initial conditions.
But there are still serious limitations that can lead to pitfalls.

George Lake reminded us that the present-day velocity dispersion in $\Omega=1$
models is too high; they are too hot.  This is an old problem, going back at
least twenty years, and I think he is correct in saying that it has not been
solved.  The standard counterargument is that the ``halos'' identified in
$N$-body simulations have smaller velocity dispersions than all mass points,
but this assumes that the internal velocity dispersions of galaxies can reach
$\sim 700\kms$, which they clearly do not.

Another well-known problem is over-merging, leading to many high-mass systems.
Ben Moore has argued that, with better resolution, substructure survives
longer.  I am not clear, however, on whether his halos converge as the
resolution length approaches zero.  In other words, is he seeing the ultimate
substructure?

The real answer is to develop computational tools that reach the required
dynamic range and have converging models.  In this regard I am very impressed
by the revolutionary method of using an adaptive multi-resolution grid reported
by Klypin and Kravtsov.  The multigrid method, invented twenty years ago by the
the Israeli mathematician Achi Brandt, is ideal for this problem.  The
innovation introduced by Klypin and Kravtsov is to populate the fine grids only
as needed, thus conserving both memory and CPU.  This allows a dynamic range
that is limited only by the number of particles, not by the size of the grid.

An extra advantage is that such an adaptive grid can equally well be used for
hydrodynamic computations with the same resolution and dynamic range.  By
contrast, the currently popular smooth particle hydrodynamics (SPH) is usually
restricted to tens of thousands of particles, compared with the millions of
grid points available in sparse multigrid hydrodynamics.  Klypin and his
colleagues are indeed developing such a hydro code, and I predict that, once
operational, it will quickly supersede SPH.

Finally, Jasjeet Bagla and Simon White pointed out the importance of
normalizing $N$-body simulations to the density of clusters at the present
epoch, which is not sensitive to the biasing of galaxy formation, unlike the
galaxy correlation function.  I agree completely and am intrigued by their
conclusion that many evolutionary effects diminish or disappear when models are
thus normalized.  I wonder, though, particularly about the evolution of the
abundance of clusters of galaxies.  It would seem to me that it has to remain a
sensitive function of the underlying cosmological model, since clusters of
galaxies correspond to the tail of the distribution of density fluctuations,
with probabilities that change rapidly as the standard deviation varies.

\section{\boldmath Explicit Versus Implicit Data Analysis}

My final remarks regard a new technological breakthrough in the extraction of
quantities of interest from noisy data.  I have in mind an integral relation
between the two of the type
\begin{equation}
D(x) = \int dy H(x,y) I(y) + N(x) \quad.  \label{eq:convolve}
\end{equation}
An example is image reconstruction, in which an image, $I$, is to be determined
after being degraded by distortion due to a point-spread function, $H$, and the
addition of noise $N$.

Explicit methods estimate the desired $I$ by operating directly on the data.
For example, if the transformation in Eq.\ (\ref{eq:convolve}) is a pure
convolution, and noise can be ignored, then $\tilde I = \tilde D/\tilde H$,
where the tilde quantities are the Fourier transforms of the equivalent spatial
functions.  Unfortunately, noise cannot be ignored, and explicit methods
amplify it, particularly at high frequencies.  In the above example, $\tilde H$
generally falls rapidly to zero at high frequency, while the noise contribution
to $\tilde D$ does not.  The result is unacceptable noise amplification.

Implicit methods attempt to overcome this difficulty by modeling $I$,
integrating it forward, and fitting the model parameters to the data.  Noise
amplification is avoided by never transforming from data space back to model
space.  The problem here is the specification of the model, which typically
uses an excessive number of parameters, resulting in over-fitting of the data.
Noise is then interpreted as real signal, which is just as bad, and sometimes
worse, than amplifying noise.

The theoretical solution to excess parameterization has been known rigorously
for thirty years, and intuitively since William of Ockham formulated his famous
razor rule in the fourteenth century.  It is to seek the solution with minimum
complexity (also known as algorithmic information content).  To state the
method in simplified terms, the idea is to use a rich language to describe the
model, e.g., by an over-complete set of basis functions, and then to seek the
solution that uses the minimum number of such basis functions, while still
adequately fitting the data.  Such models are not only efficient, they also do
the best job of separating signal and noise.  Clearly, if the data are
adequately fit with $N$ parameters, introducing another one will only serve to
fit the noise.

Unfortunately, the method of minimum complexity remained largely impractical
for a long time.  The mathematicians who invented it were quick to point out
that it was ``incomputable'', because the number of computational steps
increased too fast with the size of the dataset.  This situation has begun to
change.  A few year ago Rick Puetter and Robert Pi\~na invented a practical
application of minimum complexity to image reconstruction.  The critical new
ingredient was the realization that it was not necessary to reach the absolute
minimum of complexity.  With an intelligent search procedure, complexity is
reduced significantly after a manageable number of steps.  Beyond that, the
improvement per step declines exponentially while the computational effort
rises exponentially, so the computation is stopped.  Similar considerations
apply to many combinatorially large problems, of which the classical example
is that of the traveling salesman.

Image reconstruction in this way is known as the Pixon method.  It typically
improves resolution by a factor of a few and increases sensitivity by one to
two orders of magnitude.  The improved sensitivity comes about because of the
strong suppression of noise, which allows faint real sources to be seen where
normally they are lost in the noise.

I bring up this issue, which seemingly has little to do with the topic of this
conference, not only because we all need high-quality images, but because the
principle of minimum complexity can be generalized to a wide range of problems,
anytime the model is related to the data in a form similar to
Eq.\ (\ref{eq:convolve}).  Two examples that come to my mind immediately are
gravitational lensing, and the peculiar velocity field due to a mass
distribution.  I expect these ideas to find ample application in the coming
years.

\section*{Acknowledgments}

This work was supported in part by NASA grant AR-07551.01-96A



\end{document}